\documentclass[journal]{IEEEtran}
\usepackage{amsmath,amssymb,amsthm}
\usepackage{mathtools}
\usepackage{tikz}
\usetikzlibrary{arrows.meta,positioning,fit,backgrounds}
\usepackage[colorlinks=true,linkcolor=blue,citecolor=blue]{hyperref}

\newcommand{\rav}{\operatorname{rav}}
\newcommand{\ivec}{\vec{\iota}}

\newtheorem{identity}{Identity}
\newtheorem{definition}{Definition}

\begin{document}

\title{The Mathematical Evolution of Addressing: From Yushchenko's Address Programming Language to \emph{A Mathematics of Arrays}\\[0.35em]
{\large How Addressing Evolved into a Mathematical Theory of Indexing, Array Computation, and Machine Realization}}

\author{Lenore~M.~Mullin~and~Ga\'etan~Hains
\thanks{L. M. Mullin is Professor Emerita, College of Nanotechnology, Science, and Engineering, University at Albany, State University of New York.}
\thanks{G. Hains is Senior IEEE member, Professor of Computer Science and a member of the Laboratoire d'Algorithmique, Complexit\'e et Logique (LACL), Universit\'e Paris-Est Cr\'eteil (UPEC), France, and holds a Ph.D. in Computer Science from the University of Oxford.}}

\markboth{IEEE Annals of the History of Computing}{Mullin and Hains: The Mathematical Evolution of Addressing}

\maketitle

\begin{abstract}
This article argues that a central development in the history of computing is the mathematical evolution of addressing. During the three decades separating Kateryna Yushchenko's Address Programming Language (1958) from Lenore Mullin's \emph{A Mathematics of Arrays} (1988), addressing evolved from a machine-level mechanism for locating data into a formal theory of indexing, reduction, and machine realization, passing through Kenneth Iverson's algebra of arrays and Philip Abrams' recognition of shape as an architectural resource. Mullin's theory establishes indexing itself, via the function $\psi$, as the primitive from which array operations compose, reduce to a Denotational Normal Form, transform into an Operational Normal Form, and realize on hardware through dimension lifting. Historical claims are sourced against primary material, with inferences flagged as such; mathematical claims are stated as identities; performance claims are marked validated, theoretically established, or provisional. The result links programming languages, array mathematics, and computer architecture in one account.
\end{abstract}


\section{The Mathematical Evolution of Addressing}

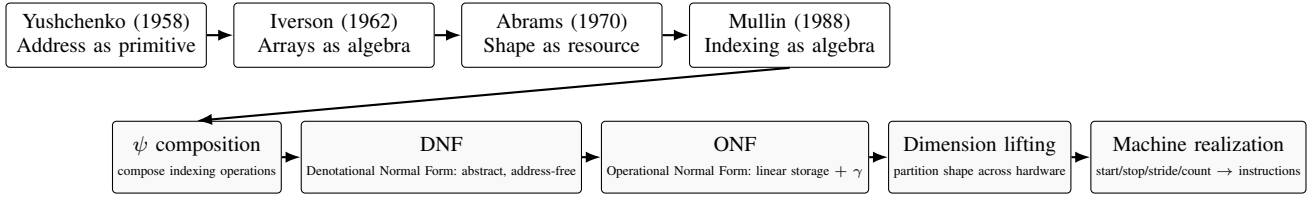
\begin{figure*}[t]
\centering
\begin{tikzpicture}[
  font=\footnotesize,
  milestone/.style={draw, rounded corners=1.5pt, align=center, minimum width=2.65cm, minimum height=0.85cm, inner sep=3pt},
  pipebox/.style={draw, rounded corners=1.5pt, align=center, minimum width=2.05cm, minimum height=0.95cm, inner sep=2pt, fill=gray!5},
  flow/.style={-{Latex[length=2mm]}, thick}
]
\node[milestone] (y) {Yushchenko (1958)\\Address as primitive};
\node[milestone, right=0.35cm of y] (i) {Iverson (1962)\\Arrays as algebra};
\node[milestone, right=0.35cm of i] (a) {Abrams (1970)\\Shape as resource};
\node[milestone, right=0.35cm of a] (m) {Mullin (1988)\\Indexing as algebra};
\draw[flow] (y) -- (i);
\draw[flow] (i) -- (a);
\draw[flow] (a) -- (m);
\node[pipebox, below=0.7cm of y, xshift=1.2cm] (pc) {$\psi$ composition\\\tiny compose indexing operations};
\node[pipebox, right=0.25cm of pc] (dnf) {DNF\\\tiny Denotational Normal Form: abstract, address-free};
\node[pipebox, right=0.25cm of dnf] (onf) {ONF\\\tiny Operational Normal Form: linear storage $+\ \gamma$};
\node[pipebox, right=0.25cm of onf] (dl) {Dimension lifting\\\tiny partition shape across hardware};
\node[pipebox, right=0.25cm of dl] (mr) {Machine realization\\\tiny start/stop/stride/count $\to$ instructions};
\draw[flow] (m.south) -- (pc.north);
\draw[flow] (pc) -- (dnf);
\draw[flow] (dnf) -- (onf);
\draw[flow] (onf) -- (dl);
\draw[flow] (dl) -- (mr);
\end{tikzpicture}
\caption{The historical closure traced in Sections II--V of this article: addressing becomes progressively abstracted from a programmable object, to an algebra of arrays, to shape as an architectural resource, and finally to a closed indexing algebra whose normal forms support systematic machine realization. The lower row expands the resulting pipeline, from composing indexing operations through the two normal forms to a realized machine instruction sequence; each stage is defined in full in Section V (DNF, ONF) and Section VI (dimension lifting, machine realization). Fig.~\ref{fig:roadmap} situates this closure within the fuller arc of formal extension, current research, and future direction.}
\label{fig:addressing-evolution}
\end{figure*}

The history of computing is often presented as the evolution of machines, programming languages, or computer architectures. This article traces it instead as the history of a single mathematical idea, \emph{addressing}, and argues that Mullin's 1988 dissertation marks the point at which indexing becomes the primitive mathematical operation from which array computation and machine realization can be systematically derived (Fig.~\ref{fig:addressing-evolution}).

Every high-level language that manipulates arrays must, at some point, answer a question that its notation may not ask explicitly: \emph{where, physically, does this element live?} Two answers to that question were given independently, on opposite sides of the Iron Curtain, within a seven-year span.

In Kyiv, from a 1957--58 seminar and a first variant jointly proposed with V.~S.~Korolyuk in 1958, Kateryna Yushchenko's Address Programming Language (Ukrainian: \textit{Adresna mova prohramuvannya}) makes address computation --- including indirect addressing and addresses of higher dimension, functionally analogous to what would later be called pointers --- the organizing primitive of the language itself. It runs on the MESM machine and, later, on Kyiv, Strela, Ural, and BESM computers, and is used operationally into the 1970s, including reportedly in the Apollo--Soyuz mission software lineage. In New York and Cambridge, in 1962, Kenneth Iverson's book \emph{A Programming Language} takes the reverse path. It is not a machine language at all; it is a notation, developed to teach and reason about algorithms on arrays, with no address model whatsoever. Iverson never conceived of shape together with a single indexing function as the basis from which every array operation could be defined; Mullin did, twenty-six years later --- where the bits live is simply not part of Iverson's theory at all.

These two traditions did not, so far as the documentary record shows, meaningfully cross-pollinate in this period; Yushchenko's work was published in Russian and Ukrainian and was, in Abrams' own later words on the general state of Soviet computer science reaching the West, essentially invisible outside the Eastern Bloc. Iverson, for his part, was building a notation for exposition, not for Soviet-style economical machine addressing. Yet both traditions converge, independently, on the same underlying problem: \emph{how does an abstract, possibly multi-dimensional index become a physical location?} This article is about the convergence, and about a single dissertation --- Philip Abrams', 1970 --- in which the Iversonian tradition is compelled, for the first time, to answer Yushchenko's question. It is also, by the end, about what it takes to answer that question not for one machine but for any machine, and not by tuning but by derivation --- which is where the historical and architectural threads of this article rejoin.

\emph{A note on names.} Both traditions produce an object commonly abbreviated ``APL'' in the secondary literature, and the coincidence is worth flagging explicitly rather than leaving implicit. Yushchenko's is the \emph{Address Programming Language}; Iverson's is, per the title of his book, \emph{A Programming Language}. They are unrelated in origin, in institutional context, and in technical character --- one is address-first, one is algebra-first --- and the only thing they share is the acronym and the underlying question this article is about.

\subsection{Scope and Method}
Three kinds of claim appear in this article, and they are held to three different standards of evidence.

\emph{Historical claims} (who did what, when, where, and under whose supervision) are held to ordinary historiographic standards: primary sources where available, explicit flagging where a draft relies on secondary sources only, and correction on the record when a detail turns out to be wrong. Section II's treatment of Yushchenko has now been checked directly against her 1963 monograph, and one such correction is recorded there, replacing a commonly repeated but inaccurate secondary-source date and attribution.

\emph{Mathematical claims} are stated as identities, in the notation of the framework being discussed, with their scope made explicit. Where MoA restates an earlier framework's problem in its own notation, this article says so explicitly rather than reading MoA notation backward into 1958 or 1962 sources that did not use it.

\emph{Architectural and performance claims} --- in particular the claim that MoA predicts, rather than merely describes, an algorithm's ideal hardware instantiation --- are the strongest and least historically settled claims in this article, and are marked accordingly: validated where supported by a published, measured benchmark, provisional where still an operating hypothesis awaiting one.

This article is, deliberately, two things at once, and it is worth saying so plainly rather than leaving a reader to infer it. Sections II--V are a history, closed as far as the documentary record presently allows: what Yushchenko, Iverson, Abrams, and Mullin actually did, checked where possible against primary sources. Sections VI--IX are something else. They argue for where the mathematical apparatus that this historical closure produced is now being extended, in research that is current, unpublished, or explicitly proposed rather than established. The two halves are kept distinct by the validated/theoretically-established/provisional labeling used throughout. Readers approaching this piece as history should nonetheless expect the second half to read more like a research agenda than a settled account, because that is what it is.

A reader interested primarily in the historical narrative, rather than the mathematical apparatus, can follow the prose in Sections II--V while treating the boxed identities and definitions as supporting detail to consult only as needed; none of the historical argument depends on verifying the algebra step by step. A reader interested primarily in the mathematics can likewise move directly to the boxed identities in Sections V--VIII, which are self-contained given the primitives defined immediately before their first use.

\section{Kateryna Yushchenko and the Address Programming Language}

\subsection{Context: MESM and the Need for a High-Level Notation}
Yushchenko (1919--2001) received her doctorate in 1950 from the Institute of Mathematics of the Ukrainian Academy of Sciences under the direction of \textbf{Boris Gnedenko}, becoming the first woman in the USSR to hold a doctorate in Physical and Mathematical Sciences in programming; she then served as Senior Researcher at the Institute (1950--57). Working with the MESM machine --- the first universally programmable electronic computer built in continental Europe --- it became apparent that direct machine-code programming did not scale to the complexity of the problems the Institute needed to solve; Yushchenko's subsequent collaboration with Glushkov on a formal mathematical description of the Kiev computer \cite{glushkov1966} shows the same addressing concerns already at work alongside machine-specific description, in a role directly comparable to what Falkoff, Iverson, and Sussenguth would later do for System/360, discussed below.

A correction to the secondary record is necessary here. Most English-language accounts (Wikipedia, popular retrospectives) date the Address Programming Language to 1955 and credit it to Yushchenko alone. Yushchenko's own 1963 monograph, \emph{Address Programming} \cite{yushchenko1963}, states the origin differently and more precisely, in its preface: the idea for a new algorithmic language arose out of a seminar on the theory of algorithms held in 1957--58, led by Prof. V.~M.~Glushkov, Prof. L.~I.~Kaluzhnin, Candidate of Physical-Mathematical Sciences V.~S.~Korolyuk, and Yushchenko herself; and the first variant of the language was proposed by Yushchenko \emph{jointly with Korolyuk} in 1958. This article follows the primary source rather than the secondary consensus: the language's origin is more accurately dated to the 1957--58 seminar and the 1958 joint first variant, not to a solo 1955 invention, though Yushchenko's sole-authored monograph three years later, and her subsequent forty-year leadership of the resulting research school, make clear that the language became substantially her own project well before 1963. L.~I.~Kaluzhnin's formalization of program--computer interfacing supplied part of the theoretical groundwork alongside this collaboration.

\subsection{The Technical Content: Addresses of Arbitrary Dimension}
Reading the primary source directly makes it possible to state the language's core primitive exactly, rather than in paraphrase. Yushchenko defines a single \emph{prime operation} (Russian: \textit{shtrikh-operatsiya}), written $'a = b$ and read ``$a$ is the address of $b$'' or ``$a$ contains $b$'': a one-argument function from a set of addresses $A$ to a set of contents $B$, required only to be single-valued and algorithmically evaluable, with no assumption that $A$ or $B$ consists of numbers. It is worth making explicit a hypothesis that Yushchenko's general definition leaves implicit but that this article's later connection to MoA (Section VIII) depends on: reading each address as, in effect, a one-dimensional array index --- an element of a set order-isomorphic to a subset of the natural numbers --- rather than an arbitrary uninterpreted symbol. Yushchenko's definition does not require this; $A$ and $B$ may be any sets whatsoever. 

When Yushchenko's quote operation is used on an address that itself contains an address, recursively --- functionally that use would later generalize to, in the Western literature, a pointer. Harold Lawson is often credited with the invention of pointers, for PL/I, in 1964. The primary source confirms this had already been done, independently, in Yushchenko's 1958 first variant of the language. 

But the specific construction that follows, addresses of higher dimension built by repeated prime application, shows that multi-dimensional arrays can be addressed with the equivalent of recursive 
quote operations if the data layout is lexicographic with respect to 
the array dimensions e.g. row-wise for two-dimensional arrays. 
Our references to Yushchenko's work do not spell-out this specific 
application of her quote operations. Yet, her work was used in 
space-related applications would naturally involve numerical computations that would normally use matrices or higher-dimensional arrays. 

So, as of today, {\bf it is a fundamental open historical conjecture:   Yushchenko and collaborators must have 
programmed multi-dimensional 
array indexing with recursive use of the quote operation. }
If the conjecture is true, they must have used repeated application of the prime operation to emulate \emph{addresses of higher dimension}: 
if $'a_1 = a_2$ and $'a_2 = b$, then $a_1$ is defined as an address of the second dimension for $b$, written $''a_1 = b$; the construction generalizes without limit to $n$-fold application, $^na = b$, with $b$ itself called the address of the zeroth dimension. 
When addresses are read this way, 
and the correspondence to $\gamma$ drawn in Section VIII is made, 
one realizes that MoA formalizes both, the multi-dimensional array 
indexing (with the $\Psi$ operator), its inverse $\gamma$ 
(which Yushchenko did not mention) {\em and } explicit shapes that 
structure the whole recursive indexing mechanism making it much 
less error prone. 

Programs written in the Address Programming Language became, notably, independent of \emph{where} in memory they were placed --- a form of position independence achieved through the addressing discipline itself. The monograph makes this explicit by distinguishing three levels (Russian: \textit{stupeni}) of the language, differing only in how ordered the underlying set of addresses is assumed to be: a fully general algorithmic level, closest in spirit to ALGOL, in which addresses may be arbitrary and sequencing is described by ordinary mathematical means (indices); a level of \emph{conditional addresses}, the one used throughout the book's own exposition, in which addresses are grouped into arithmetic sequences with an explicit successor operation, sufficient to resolve nearly all practical questions of parameter choice, memory layout, and loop construction without committing to a specific machine; and a level of \emph{concrete addresses}, in which the address set is fully ordered to match one particular target machine, so that translation from the level-two program becomes a matter of recoding rather than redesign. The same algorithm, worked as an example in the monograph, is shown written at level two and then re-expressed for the specific register and cell layout of the Ural computer, differing only in which literal addresses appear --- direct primary-source evidence for the position-independence and portability claimed for the language.

Subsequent Soviet work on the formal equivalence of address algorithms --- e.g., Korolyuk and Letichevskii's analysis of address-algorithm equivalence classes, and Malinovskii's later treatment of the same problem \cite{malinovskii1968,malinovskii1969} --- indicates the language quickly generated its own theoretical literature, independent of the Western APL tradition entirely.

\subsection{Reach and Legacy}
The Address Programming Language was implemented across first- and second-generation Soviet computers --- the monograph itself names Kiev, Strela, Ural, and BESM as machines already running address-language processing programs at the Institute of Cybernetics of the Ukrainian SSR Academy of Sciences at the time of writing --- and used through roughly two decades, in domains from aviation and ballistic trajectory calculation to, reportedly, the Apollo--Soyuz joint mission. Yushchenko went on to found what is generally regarded as the first Soviet school of theoretical programming, supervising some 47 doctoral students over her career. The fullest English-language secondary account of this period and its personalities remains Malinovsky's memoir-history \emph{Pioneers of Soviet Computing} \cite{malinovsky2010}. This section, unlike in earlier drafts of this article, is now primary-sourced rather than primary-cited, for the reason given above.

\section{Kenneth Iverson: Array Algebra Without Addresses}

\subsection{\emph{A Programming Language} (1962)}
Iverson's book \cite{iverson1962} is not, in the first instance, a language meant to be executed; it is a notation meant to be read and reasoned with, developed out of his teaching of applied mathematics at Harvard and then at IBM. Its central technical move is to give \emph{monolithic}, whole-array operations --- reduction, product, restructuring, and, among them, elementwise selection --- a compact, dimension-generalized notation, built on the idea that an array's \emph{shape} ($\rho$) is itself a first-class value that these operators generalize uniformly across, rather than being defined separately for scalars, vectors, and matrices. It is worth being precise about what this does and does not amount to: selection appears in Iverson's notation as one whole-array operator among several, not as the organizing concept of a systematic algebra. Iverson did not set out to build, and did not build, an algebra of indexing structured around shape; his notation treats an array as something one operates on as a whole, and indexing is simply one of the operations available, not the foundation the others are derived from. That reorganization --- indexing itself, understood through shape, as the thing an algebra is built around --- begins only with Mullin's Mathematics of Arrays, as Section V will show; Abrams' contribution, the subject of Section IV, is narrower: putting array shape to use as a resource for machine efficiency, without yet formalizing an indexing function of any kind. Iverson's own later reflection on the notation's purpose, in his 1979 Turing Award lecture \cite{iverson1980}, and the joint account of the language's design with Falkoff \cite{falkoff1973}, remain the standard references for the notation's intent.

\subsection{What Is Deliberately Absent --- and What It Was Used For Anyway}
It is worth stating plainly what Iverson's notation does \emph{not} contain in 1962: an address, a memory model, or any operational account of how an indexing expression is to be evaluated on physical hardware. This is not an oversight; it is the point of the exercise. Iverson's notation is, in this sense, the structural opposite of Yushchenko's language: hers begins with the address and builds a language around it; his begins with monolithic, whole-array operations, with indexing present only as one convenience among them, and leaves both the machine and the reorganization of indexing into an algebra for someone else to build.

That the notation was not itself a machine language did not stop it from being put to machine use almost immediately. In 1963--64, Falkoff, Iverson, and Sussenguth used the notation to write a complete formal description of the IBM System/360 architecture, published in the \emph{IBM Systems Journal} \cite{falkoff1964}: every functional characteristic of the machine, as seen by the programmer, stated as Iversonian array expressions rather than prose. Gerrit Blaauw, one of System/360's principal architects, is reported to have been won over by the notation through this project, going on to use it himself for teaching computer architecture after returning to the Netherlands in 1965. The causal claim that Blaauw's enthusiasm specifically motivated Hewlett-Packard to build APL directly into hardware is this article's own inference rather than one confirmed in the sources consulted, and is flagged accordingly; what is independently documented is the outcome: Hewlett-Packard's HP 3000 minicomputer, in the early 1970s, shipped with genuine APL instructions in its microcode, an implementation --- \emph{APL\textbackslash3000} --- developed with the direct participation of Larry Breed and \textbf{Philip Abrams} \cite{aplwiki3000}. That Abrams reappears here, five years after his Stanford dissertation, working on an actual hardware instruction set for APL rather than a proposed one, is a fitting bridge to the next section.

\section{Philip Abrams and \emph{An APL Machine} (1970)}

\subsection{The Problem Abrams Inherited}
By the mid-1960s, Iverson's notation had acquired implementations --- Abrams himself, with Lawrence Breed, produced in 1965 an interpreter translating Iverson's notation into IBM 7090 machine code. But an interpreter is not yet a semantics: something in the implementation has to decide, for every use of Iverson's selection operator on an array of arbitrary shape and dimension, exactly which physical storage cell is meant. Abrams' 1970 Stanford dissertation, \emph{An APL Machine} \cite{abrams1970} (Stanford Linear Accelerator Center report SLAC-R-114), directed by \textbf{Harold Stone} --- who would go on to become one of the field's most cited computer architects, author of the widely used textbook \emph{High Performance Computer Architecture} \cite{stone1993} and recipient of the IEEE Emanuel R. Piore Award and the ACM/IEEE Charles Babbage Award --- (reading committee: Bill Miller, Bill McKeeman, Ed Davidson), is the place this question is finally asked and answered rigorously, in the form of a proposed machine architecture and instruction set whose central claim is efficiency: the APL machine Abrams designs uses strictly fewer memory accesses, arithmetic operations, and temporary stores than the naive alternative, with the improvement factor scaling with operand size.

The design itself is concrete enough to be worth describing rather than only characterizing. Abrams' machine is organized around a set of machine registers together with three storage structures with distinct roles: a Value Stack (VS) holding operands and intermediate array results, a Control Stack (CS) holding the sequence of operations still to be performed, and a buffer (QS) mediating between them. Every design decision is measured against an explicit baseline Abrams calls the ``naive machine'' --- one that evaluates an APL expression the direct way, materializing every intermediate array in full before the next operation consumes it --- and the dissertation's central argument is stated as a quantitative comparison against that baseline, not as a qualitative claim of elegance. The mechanism that makes the comparison favorable is demand-driven evaluation: an operation is not performed, and the array it would produce is not materialized, until some later operation actually needs a specific element of the result, so that a selection applied to the output of an arithmetic expression can avoid computing the elements the selection would discard. This is, in the terms of this article, an operational, machine-level anticipation of what $\psi$-reduction to DNF later does algebraically --- expressing an output element as a function of input indices without materializing intermediates --- but arrived at as an implementation strategy for one architecture's efficiency, not as an algebraic identity holding for every shape and every dimension. Abrams' later work with Breed on the APL\textbackslash 3000 compiler (Section III) built on a further formalism of his own, a \emph{subscript calculus} governing how indexing expressions could be optimized \cite{aplwiki_abrams}; this, too, stops short of the general indexing algebra Section V credits to Mullin --- a calculus for optimizing subscript expressions on one implementation is not the same object as a proof that composition of indexing is, for every array, a single algebraic identity.

\subsection{Why This Dissertation Is the Hinge of the Story}
It is important to state precisely what Abrams did and did not contribute, since it is easy to overstate. Abrams introduces the idea of \emph{using array shape} as a resource for machine-level optimization: his instruction set is designed around the observation that knowing an array's shape in advance lets a machine avoid redundant memory accesses and temporary storage when evaluating Iverson's whole-array expressions. This is a machine-design insight, not an indexing function and not an algebra of indexing --- Abrams has no general indexing operator like Mullin's $\psi$, no formal object that takes an index and a shape and returns an address as a provable identity; he has a machine that happens to exploit shape efficiently for one specific architecture. Abrams' machine is, in the terms of this article, the place where shape is first put to work operationally, rather than algebraically. It does not derive an address mapping as a theorem; it builds a machine in which shape-aware addressing is realized efficiently, instruction by instruction. This is the gap this article identifies as still open in 1970: shape used as a resource for one machine's efficiency, but no indexing function, and no \emph{formal, architecture-independent algebra}, in which that use of shape is derived rather than designed. Closing that gap --- inventing the indexing function itself, $\psi$, and the algebra built around it --- is the project of the Mathematics of Arrays.

\section{The Mathematics of Arrays: Closing the Gap}

\subsection{Restating the Problem in One Notation}
The unresolved question inherited from Sections II--IV can now be posed precisely, using Iverson-derived MoA notation. Let $\xi$ be an array, let $\rho\xi$ denote its shape, let $\iota$ denote the index generator, and let $\psi$ denote the indexing function. Types make the relationships among these primitives explicit. $\rho: \mathrm{Array}[\tau] \to \mathrm{Vector}[\mathbb{N}]$ takes an array of any element type $\tau$ to its shape, a vector of natural numbers, one entry per dimension. $\iota$ has two types, corresponding to the two cases distinguished in Section V.E below: $\iota: \mathbb{N} \to \mathrm{Vector}[\mathbb{N}]$ for a scalar argument, returning a flat range, and $\iota: \mathrm{Vector}[\mathbb{N}] \to \mathrm{Array}[\mathrm{Vector}[\mathbb{N}]]$ for a vector (shape) argument, returning an array of index vectors rather than a single vector. $\psi: \mathrm{Vector}[\mathbb{N}] \times \mathrm{Array}[\tau] \to \tau$ takes an index vector together with an array of element type $\tau$ and returns a single element of that type; applied to a partially-bound index vector it returns an array of type $\tau$ rather than a scalar, as Section V.B makes precise. The first identity below is the base case --- indexing $\xi$ by the full generator of its own index space returns $\xi$ itself, i.e.\ indexing is faithful to shape:

\begin{identity}[Index-space identity]
\[
(\iota\,\rho\,\xi)\;\psi\;\xi \;\equiv\; \xi .
\]
\end{identity}

This says nothing yet about \emph{where} the elements of $\xi$ live; it is a statement purely in index space --- the formal indexing function $\psi$ itself, together with this identity, is MoA's invention, with no counterpart in either Iverson's monolithic whole-array notation or Abrams' shape-aware machine design. It is worth being precise about the strength of this claim: the identity is not merely absent from APL as a matter of historical record; it is not a provable law in APL, or in any other existing array language or library, today. APL, NumPy, MATLAB, and their successors all provide operators that \emph{look like} $\iota$, $\rho$, and indexing, and a user can certainly write an expression that evaluates, for a particular $\xi$, to something extensionally equal to $\xi$ --- but none of these languages states or proves, as a general algebraic law holding for every shape and every dimension, that indexing an array by the full generator of its own index space returns that array. The identity is a theorem of MoA specifically, built on the closure Mullin put on an algebra of arrays constructed from exactly three primitives --- the indexing function $\psi$, the shape function $\rho$, and the index generator $\iota$ --- not an incidental fact that happens to be true of arrays in general and could equally well be stated in any of these other languages. The second identity is where Yushchenko's question --- \emph{where, physically?} --- re-enters, algebraically rather than operationally:

\begin{identity}[Address-mapping identity]
\[
\ivec\;\psi\;\xi \;\equiv\; (\rav\,\xi)\big[\gamma(\ivec ; \rho\,\xi)\big].
\]
\end{identity}

Here $\rav\,\xi$ is the ravel of $\xi$ --- its canonical flattening into one-dimensional storage order (the Operational Normal Form, ONF, target of the DNF--ONF pipeline) --- and $\gamma$ is the mapping function that takes an arbitrary index vector $\ivec$, together with the shape $\rho\xi$ against which it is being interpreted, and returns a single linear address into that flattened storage. In types: $\rav: \mathrm{Array}[\tau] \to \mathrm{Vector}[\tau]$ takes an array of any shape to a vector holding the same elements in a single fixed linear order, and $\gamma: \mathrm{Vector}[\mathbb{N}] \times \mathrm{Vector}[\mathbb{N}] \to \mathbb{N}$ takes an index vector together with a shape (both vectors of natural numbers) and returns a single natural number, the offset into $\rav\,\xi$ at which the indexed element is stored. The canonical order $\rav$ flattens to was originally assumed to be row-major, matching Iverson's own convention; it need not be. $\rav$, reshape, and $\gamma$ are all parameterized by whatever linear ordering the target storage actually uses --- row-major, column-major, or any other total order on the index space --- and each is defined relative to that choice rather than committed to row-major specifically, which is what allows dimension lifting (Section VI) to assign a different, locally contiguous ordering at each level of a memory hierarchy without contradicting Identity 2 at the level below it.

\subsection{Scalars, Sub-Arrays, and Shape as Retained Hardware Context}
Identity 2 is stated for an arbitrary index vector $\ivec$. The word ``arbitrary'' is doing real work here, and it is worth spelling out why. A fully-bound index vector has one entry per dimension of $\rho\xi$; $\gamma$ maps it to the single offset of one scalar element. A partially-bound index vector leaves some dimensions fixed and others free to range over their extent; $\gamma$ maps it instead to the whole set of offsets swept out by a sub-array (a section, or slice) of $\xi$. A scalar access and a sub-array access are therefore the same operation, $\gamma$, applied to index vectors of different dimensionalities --- neither needs a separate slicing rule bolted on afterward. This is what lets one identity, rather than one rule for elements and a second rule for sections, carry the address-mapping burden through both dimension lifting (Section VI) and indirection (Section VIII).

Identity 1 also admits a second, fully concrete restatement. Here a correction is needed to a natural first guess at what that restatement should say. One might write the right-hand side as a physical rebuild: ravel $\xi$ down to a flat run of storage, then \emph{reshape} that run back up into a multi-dimensional structure. But this is not what happens on the hardware (ONF) side, and it is not what should be claimed. Storage stays linear throughout; nothing is ever physically restructured into multi-dimensional form. What happens instead is simpler: the shape $\rho\xi$ is \emph{retained}, as context, alongside the flat storage. $\gamma$ then computes the address of any element on the fly, at access time, directly from that retained shape. Nothing is rebuilt into $\xi$'s original rectangular layout; the shape is just kept on hand. The correct restatement is therefore stated directly in terms of $\gamma$, not reshape:
\begin{identity}[Hardware-side restatement of Identity 1]
\[
(\iota\,\rho\,\xi)\;\psi\;\xi \;\equiv\; (\rav\,\xi)\big[\gamma(\iota\,\rho\,\xi ; \rho\,\xi)\big],
\]
\end{identity}
Here $\gamma$ is applied elementwise, once per index vector in the array $\iota\,\rho\xi$. Each application uses the shape $\rho\xi$, carried alongside $\rav\,\xi$ as $\gamma$'s second argument, to compute that one element's address individually and on demand. Nothing about $\xi$'s original structure is rebuilt in storage. $\rho\xi$ is kept in the hardware environment for one reason only: so that addressing --- linear-layout access for an arbitrary index --- can be computed at the moment it is needed. This identity is, in fact, simply Identity 2 specialized to one case: where $\ivec$ is the full index generator $\iota\,\rho\xi$, rather than a single index vector. It is also the correct account of what dimension lifting (Section VI) carries forward to each level of the memory hierarchy. What moves down through the hierarchy is not a reshaped copy of the data --- it is the shape itself, retained as the information $\gamma$ needs to keep computing addresses on the fly.

\subsection{The Full Primitive Set Linking DNF to ONF}
Identities 1 through 3 have, in fact, already been using two further primitives silently, on top of $\psi$ and $\gamma$: the shape operator $\rho$ and the ravel operator $\rav$ (both, like $\iota$ itself, original APL\textbackslash 360 primitives due to Falkoff and Iverson \cite{falkoff1973}, not new to MoA), together with bracket selection $[\,]$. $\rho\xi$ is what $\iota$ is applied to in Identity 1, what $\gamma$ takes as its second argument in Identity 2, and it is $\rho\xi$, retained as context, that Identity 3 identified as the information carried across the DNF/ONF boundary. $\rav\,\xi$ is what makes the ONF side of every identity in this article concrete in the first place: it is the operator that takes $\xi$, however many dimensions it has, down to the single flat run of storage that $\gamma$'s addresses actually index into --- without $\rav$, $\gamma$ would have nothing linear to address. Bracket selection $[\,]$ is what actually reaches into $\rav\,\xi$ once $\gamma$ has produced an address, and reaches into an index array once $\iota$ or $\gamma$ has produced an index into it, as in Section VIII's indirection identity. What is new is not $\rho$, $\iota$, or $\rav$ themselves, but the fact that $\psi$ and $\gamma$ --- MoA's actual inventions --- give this pre-existing Iversonian vocabulary a job it never had: together they are what actually links the DNF to the ONF, in a sense Iverson's own algebra never assembled them to do. The DNF is reached once $\rho$ has produced a shape and $\iota$ has produced an index generator over it, and the ONF is reached once $\rav$ has flattened $\xi$ to linear storage, $\gamma$ has consumed the shape to compute an address into it, and $[\,]$ has performed the resulting access. Put plainly, the algebra is, at its core, \emph{indexing with shapes}: $\psi$ and $\gamma$ exist solely to relate an array's shape --- built from Iverson's own $\rho$, $\iota$, and $\rav$ --- to the index used to access it, from the moment that index is first generated in abstract coordinate space to the moment it becomes a concrete offset into linear storage.

It is worth being explicit about the scope of $\gamma$'s codomain, since the natural comparison to C-style pointers can otherwise mislead. $\gamma$ returns a natural number: an offset, counted in elements, into the linear order $\rav\,\xi$ imposes on $\xi$. This is a genuinely narrower space than the one C pointers occupy. A C pointer is, in general, an arbitrary bit pattern in a machine word --- large enough to encode not just an offset but tagged metadata or an address outside any single array's own extent entirely; its type is closer to an arbitrary element of $\{0,1\}^w$ for some word width $w$ than to $\mathbb{N}$. MoA's $\gamma$ deliberately does not attempt this generality: it is defined only over the index space of one array's own shape, and its output is meaningful only relative to that array's own $\rav$. Realizing a $\gamma$-computed offset as an actual pointer --- adding it to a base address, applying a stride in bytes rather than elements, respecting a target's specific pointer representation --- is exactly the translation step described in Section VI.D: the ONF's natural-number offsets are what a translator maps onto whatever pointer representation the target language or hardware actually uses, and that mapping, not $\gamma$ itself, is where C-pointer-style generality enters.

That $\rho$ can be applied to its own output --- shape is itself an array, and so has a shape of its own --- gives the cleanest illustration of why $\rho$ and $[\,]$ belong in the primitive set explicitly rather than left implicit. Let $\delta\xi$ denote the dimension of $\xi$. Ordinarily treated as separate metadata supplied by an implementation, $\delta\xi$ is instead derivable from $\rho$ and $[\,]$ alone:
\begin{identity}[Dimension identity]
\[
\delta\xi \;\equiv\; \big(\rho(\rho\,\xi)\big)[0].
\]
\end{identity}
$\rho\xi$ is a vector with one entry per dimension of $\xi$; taking its own shape, $\rho(\rho\xi)$, yields a one-element vector holding exactly the dimension of $\xi$, since the shape of any vector is a one-element vector containing that vector's length; bracket-selecting that vector's sole entry, $[0]$, yields $\delta\xi$ itself, as a scalar. No operation outside $\rho$ and $[\,]$ is needed. The bracket notation here is itself only shorthand for $\psi$: the same identity could equally be written
\[
\delta\xi \;\equiv\; \langle 0 \rangle \; \psi \; \big(\rho(\rho\,\xi)\big),
\]
making explicit that $[\,]$ never names an operation beyond $\psi$ itself. The two notations are not, however, interchangeable at full generality: bracket selection $X[i]$ is used in this article specifically when $X$ is a vector and $i$ a scalar index into it, whereas $\psi$ is the fully general operator, applying an index vector to an array of any shape and dimension. $X[i]$ is therefore shorthand for the vector case only, $\langle i \rangle \, \psi \, X$, and is retained purely for readability in that one case, not as a general substitute for $\psi$ at arbitrary dimension. The two notations also divide, in practice, along the DNF/ONF boundary itself, though $[\,]$ remains defined for vectors on either side of it: $[\,]$ is the notation this article reaches for in the ONF world, where what is being described is an offset from an address into linear storage --- exactly the role $\rav\,\xi[\gamma(\ivec;\rho\xi)]$ gives it in Identity 2 --- while $\psi$ is the notation used in the DNF world, where the argument is still an array and no address has yet been computed. The point of writing the dimension identity both ways is to show that indexing a scalar --- selecting the single entry of the one-element vector $\rho(\rho\xi)$ --- is the same operation, $\psi$, whether one is working in the DNF world (an abstract coordinate, not yet an address) or the ONF world (a concrete offset into linear storage): $\psi$ does not change character depending on which side of the DNF/ONF boundary it is applied on, and a scalar is indexed identically in either regime, even though $[\,]$ is the notation this article would ordinarily reach for on the ONF side. This is the same discipline this article follows throughout: what looks like an independent piece of information --- here, dimension itself --- is in fact derivable, on demand, from the same small primitive set that already carries every array expression from abstract index to physical address.

\subsection{A Worked Example}
The identities above are stated once, for an arbitrary array $\xi$; it is worth fixing one small, concrete $\xi$ and checking each identity against it directly, since the algebra is easy to lose behind the notation on a first reading. Take
\[
\xi \;=\; \begin{pmatrix} 10 & 11 & 12 \\ 13 & 14 & 15 \end{pmatrix},
\]
a $2\times 3$ array, so $\rho\xi = \langle 2, 3\rangle$ and $\delta\xi = 2$. Fix row-major storage, so
\[
\rav\,\xi \;=\; \langle 10, 11, 12, 13, 14, 15 \rangle,
\]
elements listed row by row. This single array is used to check Identities 1 through 4 in turn, and is carried forward into the worked example in Section VI.

\emph{Identity 1.} $\iota\,\rho\xi$ generates every index vector into a $2\times 3$ array: $\langle 0,0\rangle, \langle 0,1\rangle, \langle 0,2\rangle, \langle 1,0\rangle, \langle 1,1\rangle, \langle 1,2\rangle$, arranged themselves into a $2\times 3$ array matching $\rho\xi$. Selecting $\xi$ at each of its own index vectors, in place, reproduces $\xi$ exactly --- $(\iota\,\rho\xi)\,\psi\,\xi \equiv \xi$ --- which is only interesting to check at all because the claim is that this holds for \emph{every} array, not because the arithmetic is hard.

\emph{Identity 2.} Take the index vector $\vec\iota = \langle 1, 2\rangle$, selecting row $1$, column $2$ of $\xi$: directly, $\xi[1,2] = 15$. Via the identity, $\gamma(\langle 1,2\rangle ; \langle 2,3\rangle)$ computes the row-major offset $1 \times 3 + 2 = 5$, and $(\rav\,\xi)[5] = 15$ --- the sixth entry (indexing from $0$) of $\langle 10,11,12,13,14,15\rangle$. Both routes agree, as the identity requires.

\emph{Identity 3.} Applying $\gamma$ elementwise across all six index vectors of $\iota\,\rho\xi$, in the order those index vectors are generated, gives the address sequence $\langle 0,1,2,3,4,5\rangle$ for this particular row-major $\xi$ --- which is the identity permutation only because the index generator and the storage order happen to agree here. Switching to column-major storage for the same $\xi$ would leave every element value unchanged but permute this address sequence to $\langle 0,3,1,4,2,5\rangle$ instead, illustrating concretely the point made above Identity 2: $\gamma$ is parameterized by the storage order, not committed to row-major.

\emph{Identity 4.} $\rho\xi = \langle 2,3\rangle$; $\rho(\rho\xi) = \rho\langle 2,3\rangle = \langle 2 \rangle$, a one-element vector, since $\langle 2,3\rangle$ itself has length $2$; and $\big(\rho(\rho\xi)\big)[0] = 2$, matching $\delta\xi = 2$ as required.

\subsection{Reading the Identity Historically}
Identity 2 is, in effect, the formal object that Abrams' machine exploited only as a resource for one architecture's efficiency in 1970, and that Yushchenko's language took as an unanalyzed primitive from 1958. MoA's contribution is to make $\gamma$ itself a derived function of the algebra --- subject to the same dimension-generalized DNF-to-ONF reduction pipeline as every other MoA expression --- so that the address computation for a given array expression is not chosen by an implementer, nor built into a fixed machine architecture, but \emph{derived, once, algebraically, and proven equivalent to the direct semantics of $\psi$ for every shape and every dimension.} This is the sense in which MoA can be read as the synthesis this article has been building toward: Yushchenko's insistence that the address is the thing to reason about, combined with Mullin's invention of indexing itself as a uniform algebra over shape --- present in neither Iverson's dimension-generalized whole-array notation nor Abrams' shape-aware machine design --- resolved by giving $\gamma$ itself the status of a provable identity rather than either a raw language primitive or a hand-built machine.
Identity 2 is, in effect, the formal object that Abrams' machine exploited only as a resource for one architecture's efficiency in 1970, and that Yushchenko's language took as an unanalyzed primitive from 1958. MoA's contribution is to make $\gamma$ itself a derived function of the algebra --- subject to the same dimension-generalized DNF-to-ONF reduction pipeline as every other MoA expression --- so that the address computation for a given array expression is not chosen by an implementer, nor built into a fixed machine architecture, but \emph{derived, once, algebraically, and proven equivalent to the direct semantics of $\psi$ for every shape and every dimension.} This is the sense in which MoA can be read as the synthesis this article has been building toward: Yushchenko's insistence that the address is the thing to reason about, combined with Mullin's invention of indexing itself as a uniform algebra over shape --- present in neither Iverson's dimension-generalized whole-array notation nor Abrams' shape-aware machine design --- resolved by giving $\gamma$ itself the status of a provable identity rather than either a raw language primitive or a hand-built machine.

\subsection{Closure: Mullin, \emph{A Mathematics of Arrays} (1988)}
The three-stage attribution this article has been assembling can now be stated plainly, and it is worth stating as a claim about closure, not merely influence. Iverson (1962) supplies a dimension-generalized algebra of arrays: monolithic, whole-array operations, in which $\iota$ is defined only for a scalar argument, with a one-element vector treated as the same argument as the scalar under APL's general scalar-extension convention, and never given its own definition for a vector of arbitrary length; selection by index is present only as one operator among several. There is no indexing function in this algebra, and no algebra of indexing composition to speak of --- because indexing was never the thing being organized. Abrams (1970) introduces the idea of using array shape as a resource: his machine exploits foreknowledge of shape to cut memory accesses and temporary storage for one specific architecture, and this shape-aware optimization, generalized well beyond Abrams' own machine, went on to guide the design of Fortran~90, MATLAB, NumPy/SciPy, Julia, and their successors --- but Abrams has no indexing function either, only a machine that uses shape efficiently.

It is Mullin's 1988 Syracuse PhD dissertation, \emph{A Mathematics of Arrays} \cite{mullin1988} --- a single-author dissertation, and the closure credited in this section is Mullin's alone, prior to and independent of any later collaborative work --- that closes this open question, and it is worth being exact about what closing it required inventing versus what it repurposed. $\rho$ (shape), $\iota$ (the index generator, APL's Iota), and $\rav$ (ravel, APL's unary comma) are not new: all three are original APL\textbackslash 360 primitives, due to Falkoff and Iverson \cite{falkoff1973}, and Iverson's own algebra already used them --- but only for monolithic whole-array operations, never assembled into an algebra of indexing. What MoA actually invents are $\psi$, the generalized indexing function, and $\gamma$, the address-mapping function, together with the identities that weld all five independent primitives --- $\psi$, $\rho$, $\iota$, $\rav$, and $\gamma$ --- into a single closed algebra in which composition of indexing is a provable identity rather than an implementation choice. Bracket selection $[\,]$, introduced alongside them in Section V.C, is not a sixth independent primitive; it is shorthand for $\psi$ in the vector case, as already shown there. Neither Iverson's algebra of arrays nor Abrams' shape-aware machine contains $\psi$ or $\gamma$, or anything equivalent to them; MoA's contribution is not a new vocabulary but a new structure imposed on an existing one, in which $\rho$, $\iota$, and $\rav$ retain their original Iversonian meaning while $\psi$ and $\gamma$ put them to a use Iverson's own algebra never assembled them for. This is not a small difference in degree: APL's own primitive functions --- $+$, $\lceil$, reduction, product, and the rest --- were never themselves defined in terms of $\psi$ and shape; each was specified on its own terms, with dimension and conformability rules stated case by case rather than derived from a single underlying indexing algebra. Critically, \emph{any} function over arrays can be defined in terms of shapes and $\psi$ alone in MoA, which is a claim APL's own definitional practice neither makes nor supports.

Inventing $\psi$ also means resolving an anomaly original to Iverson himself, not merely to later dialects. Iverson defined $\iota$ only for a scalar argument $n$ --- $\iota\,n$ generates the flat range $\langle 0,1,\ldots,n-1\rangle$ --- and, under APL's general scalar-extension convention, treated a one-element vector $\langle n\rangle$ as the same argument as the scalar $n$, so that $\iota\,n$ and $\iota\,\langle n\rangle$ collapse to one case rather than being distinguished. Iverson never defined $\iota$ for a vector argument of arbitrary length, the case a shape $\rho\xi$ actually is once $\xi$ has dimension greater than one. MoA removes this anomaly rather than inheriting it: it defines $\iota$ for a vector argument in general, of any length, yielding the multi-dimensional generator $\iota\,\rho\xi$ used throughout Identity 1 and the identities that follow it --- an array of index vectors, one per dimension of $\rho\xi$ --- as a case genuinely distinct from, and never collapsed into, the scalar case. What Iverson left as one undifferentiated case, MoA defines as two.

Composition of indexing is no longer a machine-level folding heuristic; it is the DNF-to-ONF reduction, a derivation that any composed chain of $\psi$-expressions over arbitrary shape and dimension is provably equivalent to a single, address-optimal normal form. This is the sense in which \emph{A Mathematics of Arrays} should be read as the point of closure for the historical arc traced in this article: Yushchenko opens the question of address as primitive (1958); Iverson supplies a dimension-generalized algebra of whole-array operations, without addresses, without an indexing function, and without a settled definition of $\iota$ beyond a scalar argument (1962); Abrams introduces the use of array shape as a machine-level resource, still without an indexing function of any kind (1970); Mullin invents the indexing function itself and closes the algebra formally, for every machine, giving $\iota$ its own precise vector-argument definition along the way (1988).

\section{Dimension Lifting: From Optimal Computation to Predicted Instantiation}

\subsection{The Problem Beyond Composition}
Closing the composition-of-indexing question --- the achievement credited to Mullin's 1988 dissertation above --- settles what an optimal \emph{abstract} computation looks like: the DNF produced by $\psi$-reduction is architecture-independent, eliminates intermediate arrays, and expresses every output element as a closed-form function of input indices alone. This is worth connecting explicitly back to Yushchenko, rather than left as a parallel that only Iverson and Abrams were credited with in Sections III--V. Yushchenko's own intermediate level --- \emph{conditional addresses}, Section II.B --- already occupies exactly this position: addresses grouped into arithmetic sequences with an explicit successor operation, sufficient to resolve memory layout and loop construction without yet committing to any specific machine's concrete address set. The DNF's flat index enumeration, $\iota\,\rho\xi$ ravel-ordered, is the same idea made fully general: a linearly-ordered enumeration of abstract positions, prior to $\gamma$'s commitment to a concrete storage layout, exactly as Yushchenko's conditional addresses stood prior to her own level-three commitment to a specific machine's register and cell layout. The historical sequence is therefore not only Yushchenko-address, Iverson-algebra, Abrams-shape, Mullin-closure; it is, at this specific point, Yushchenko-conditional-address, MoA-DNF --- the same abstraction level, arrived at independently, thirty years apart. But an abstract optimum is not yet a placement on physical hardware with a memory hierarchy --- registers, cache lines, NUMA domains, GPU warps, multiple sockets. A further mechanism is needed to take the DNF's shape information and distribute it across such a hierarchy without leaving the algebra --- that is, without falling back to the kind of architecture-specific hand-tuning that Abrams' machine represented for one design in 1970. This is the role of \emph{dimension lifting}.

\begin{definition}[Dimension lifting]
Dimension lifting \cite{grout2022,gpu2023} --- anticipated in Mullin and Raynolds' earlier work connecting the hardware/software boundary algebraically \cite{mullin2008} and extending MoA's composition of tensor and array operations to explicit hardware shapes \cite{mullin2009} --- is the process by which the shape $\rho\xi$ of a $\psi$-reduced (DNF) array expression is partitioned into a tuple of sub-shapes, each assigned to a component of a target architecture --- a vector register, a cache line, a processor, a NUMA socket, a GPU warp --- with every such component itself modeled, uniformly and in the same Cartesian formalism, as an array of known shape. The partitioning increases the effective dimension of the computation: a 2-dimensional algorithm, already optimized through $\psi$-reduction, may become 3-dimensional once a processor dimension and a register dimension are added, or $n$-dimensional for $n$ chosen hardware resources.
\end{definition}

The key move is conceptual rather than merely technical: hardware and data are represented in the \emph{same} algebra. A cache line is not an external constraint bolted onto the computation after the fact; it is a shape, and shapes compose exactly as data shapes compose. This is what allows the mapping from the optimized DNF to the ONF to remain a derivation --- each level of the memory hierarchy contributes its own $\gamma$, and the full address decomposes as a composition of per-level $\gamma$'s, $\gamma = \gamma_1 \circ \gamma_2 \circ \cdots \circ \gamma_k$, one per level of the partition, exactly as Identity 2 composes for a single flat address space.

\subsection{The Worked Example, Continued}
Take the same $\xi$ from Section V.C, the $2\times 3$ array with $\rho\xi = \langle 2,3\rangle$, and suppose the target hardware offers two memory banks, addressed independently. Dimension lifting partitions $\xi$'s row dimension across them: bank $0$ holds row $0$ of $\xi$, bank $1$ holds row $1$. This is exactly the shape-partitioning Definition 1 describes --- $\rho\xi = \langle 2,3\rangle$ is split into a bank dimension of extent $2$ and a within-bank dimension of extent $3$ --- and it raises the effective dimension of the computation from $2$ to $3$: a bank index, a row-within-bank index (here always $0$, since each bank holds exactly one row), and a column index.

Address computation now factors exactly as $\gamma = \gamma_1 \circ \gamma_2$ describes in general. $\gamma_1$ takes the row component of an index vector and selects a bank: row $0 \to$ bank $0$, row $1 \to$ bank $1$. $\gamma_2$ then computes an offset \emph{within} that bank's own local storage, which holds only three elements, using the bank's own local $\rav$: bank $0$'s local storage is $\langle 10,11,12\rangle$, bank $1$'s is $\langle 13,14,15\rangle$, and each is addressed $0$--$2$ independently of the other. To retrieve $\xi[1,2] = 15$ under this partitioning: $\gamma_1$ sends row $1$ to bank $1$; $\gamma_2$ sends column $2$ to local offset $2$ within bank $1$'s storage; bank $1$'s storage at local offset $2$ is $15$. This is the same value Identity 2 computed directly against the single flat $\rav\,\xi$ in Section V.C, reached instead by composing two smaller, bank-local address computations --- exactly the sense in which dimension lifting is a derivation from the same algebra, not a separate, hand-tuned partitioning scheme layered on top of it.

\subsection{Why This Licenses a Predictive, Not Merely Descriptive, Claim}
A framework that only \emph{describes} an already-built implementation post hoc is not doing the same work as one that \emph{derives} an implementation from the algebra before any implementation exists. Because dimension lifting partitions shapes algebraically --- as a function of the DNF's shape and the target hardware's shape, both expressed in the identical formalism --- the mapping from ``algorithm expressed as an array expression'' to ``instantiation on a given memory hierarchy'' is not a search over a space of candidate implementations, nor a heuristic tuned empirically per platform. It is a derivation: given the shape of the problem and the shape of the hardware, the partitioning of $\rho\xi$ that yields contiguous access at every level is determined, not discovered. This is the basis for the claim --- validated, not merely asserted, in the recent cache-blocking and GPU energy-efficiency work built on this framework \cite{thomas2021,gpu2023} --- that if an algorithm is formulated as an array (tensor) expression in MoA, the framework does not stop at optimizing the abstract computation: it can be used to predict, and to construct, that algorithm's ideal instantiation on a specified target architecture, because the same $\psi$/$\iota$/$\gamma$ apparatus that proves the DNF optimal is reapplied, without modification, to the partitioning of the hardware's own shape.

The precision of that partitioning is not fixed once and for all by the algebra alone; it is bounded by how much is actually known, and controllable, about the target machine's performance characteristics. Current research extends this point directly: the more access a derivation has to a machine's real performance behavior --- measured bandwidths, latencies, and achievable throughput, rather than nominal or vendor-stated figures --- and the more that behavior can be controlled or held fixed during measurement, the tighter the upper bounds that can be derived on achievable performance, and the more precisely dimension lifting can partition $\rho\xi$ against those bounds \cite{cost2026}. Dimension lifting's optimality claim, in other words, is optimal relative to the accuracy of the hardware model it is given; improving access to and control over machine performance is not a separate empirical concern running alongside the algebra, but a direct input to how tightly the algebra can dimension-lift.

\subsection{Relation to the Current Benchmarking Program}
This claim divides, per the Scope and Method statement above, into a validated part, a theoretically established part, and a still-empirical part. \emph{Validated}: MoA-derived cache-blocking for GEMM has been shown, in Thomas, Mullin, and \'Swirydowicz's measured benchmarks, to outperform established DGEMM libraries \cite{thomas2021}, and MoA-derived data/hardware-shape partitioning for GPU energy efficiency has likewise been published with measured performance comparisons against classically hand-tuned implementations \cite{gpu2023}; in both cases the derivation --- not a search over blocking parameters --- is what produces the winning configuration. \emph{Theoretically established}: for transformer attention specifically, the MoA-derived kernel is not merely conjectured to be data-movement optimal; a lower-bound theorem is proved directly, showing that any correct implementation of scaled dot-product attention must read and write at least the amount of data the MoA-derived kernel actually reads and writes, so the derived kernel meets a proven lower bound rather than merely improving on a baseline \cite{attention2026}. \emph{Still empirical}: what remains open is not the theoretical optimality claim but its \emph{measured} realization on specific hardware --- the object of the five-implementation matrix-multiplication benchmark on Apple M1 Pro hardware, its accompanying unified predictive performance model, and the ACCESS Accelerate allocation request to run MoA-derived transformer kernels experimentally on shared HPC hardware for the first time. Whether the measured performance tracks the proven bound as closely as the abstract theory predicts is precisely what the current benchmarking program is designed to find out, and this article will report the result, whichever way it comes out, rather than assuming it in advance.

\subsection{From ONF to Instructions: A Translator, Not a Compiler}
It is worth being precise about what kind of tool is actually needed once dimension lifting has produced the ONF, since the answer is more modest than the word ``compiler'' usually implies. The ONF describes every array operation in terms of four primitives --- start, stop, stride, and count --- a universal machine abstraction \cite{grout2018,grout2022} that already specifies, for a given target architecture, exactly which memory locations are touched, in what order, and with what step. Because dimension lifting has already assigned shapes to hardware components before this point, the ONF is simultaneously optimal and architecture-specific by the time it exists at all. What stands between the ONF and executable code is therefore not optimization in the usual compiler sense --- discovering a loop order, choosing a blocking factor, deciding whether to vectorize --- but translation: a one-to-one mapping from ONF start/stop/stride/count expressions to a target's native instructions, compiler pragmas (OpenMP, OpenACC), or intrinsics --- or, on the hardware-description side, to VHDL. The same start/stop/stride/count description that a translator maps to software instructions can instead be mapped to a synthesizable VHDL design, since both are just target-specific realizations of the same ONF; Grout and Mullin have demonstrated this directly, taking a Python program formulated in MoA through DNF and ONF and building verified VHDL hardware from it, with no separate hardware-design step distinct from the translation itself \cite{grout2018,grout2022}. A conventional optimizing compiler for array code carries the burden of proving, or simply trusting, that a chosen transformation preserves the source program's meaning; an ONF translator carries no such burden, because correctness was already established once, algebraically, at the DNF-to-ONF derivation, and the translation step need only preserve the one-to-one correspondence rather than reconstruct or re-verify it. This is the same guarantee invoked below for the Python-to-C pipeline in Section IX --- a translator standing between a provably optimal normal form and its physical execution, not a compiler standing between a specification and a hoped-for optimization.

\section{Confluence: Church--Rosser and the Case for a Unified Formal Functional Array Paradigm}

\subsection{Why Confluence Matters, Not Just Efficiency}
Sections V and VI made two claims that this section now grounds formally rather than leaving as assertions: that a composed chain of $\psi$-expressions reduces, provably, to a single DNF regardless of how the composition happens to be written; and that dimension lifting can proceed level by level without the order of partitioning affecting correctness. Both claims are, precisely, claims of \emph{confluence}: that a rewriting system reaches the same result no matter which order its rewrite rules are applied in. This is exactly the property Church and Rosser proved for the $\lambda$-calculus in 1936, and it is the property that gives functional programming, as a discipline, its central guarantee --- that a program's meaning does not depend on an evaluator's scheduling choices.

\begin{definition}[Church--Rosser / confluence]
A rewriting system has the Church--Rosser property if, whenever an expression $e$ reduces (in zero or more steps) to both $e_1$ and $e_2$, there exists some $e'$ to which both $e_1$ and $e_2$ further reduce. Consequently, if $e$ has a normal form, that normal form is unique, independent of the order in which reduction rules were applied.
\end{definition}

It is worth being precise about what this definition does and does not guarantee, since the qualifier ``if $e$ has a normal form'' is doing real work and is easy to read past. Confluence guarantees uniqueness of the DNF conditional on one existing; it says nothing about whether $\psi$-reduction of an arbitrary MoA expression is guaranteed to \emph{terminate} in the first place. Termination (strong normalization) is a separate property from confluence, and a rewriting system can be confluent without being terminating.

Stating either question precisely, however, requires one further step that this article --- and, to its knowledge, the MoA literature more broadly --- has left implicit up to this point: obtaining an actual rewriting system from MoA's identities requires \emph{orienting} them. Identities 1 through 5 are stated as equivalences, $\equiv$, true and usable in either direction; a rewriting system, by contrast, requires each identity to be read in one direction only, as a rule that replaces a left-hand-side pattern with a right-hand-side term and never the reverse. For $\psi$-reduction, the natural orientation reads every identity left to right --- an indexing expression rewrites to its DNF, never the other way --- but this orientation has not been stated explicitly anywhere in the MoA literature to date, this article included. Making it explicit is a precondition, not an afterthought: only once the identities are oriented into directed rewrite rules do confluence and termination become precisely posed questions, open to the standard term-rewriting toolkit --- critical-pair analysis for confluence, a well-founded reduction ordering for termination --- rather than properties asserted by analogy to the $\lambda$-calculus without a rewriting system in hand to check them against. This article treats orienting the MoA identities as itself the open problem prior to, and a precondition for, actually proving rather than citing the confluence and termination results discussed below.

This article has not located, and does not claim to supply, a termination proof for $\psi$-reduction distinct from the confluence property discussed below; the practical experience across the compiler and benchmarking work cited throughout this article is that $\psi$-reduction of the array expressions actually encountered does terminate, but that is an empirical observation across worked examples, not a general theorem, and stating it as a theorem would overreach the Scope and Method commitments of this article. It is worth narrowing the open question rather than leaving it maximally general: every application surveyed in this article --- the GEMM and GPU benchmarks, the attention kernels, the Fortran and Python compilers, the image-processing pipeline --- is a primitive-recursive function, or a still less expressive one, and primitive-recursive functions terminate by construction, independent of any confluence argument. The open question this article carries forward is therefore not whether $\psi$-reduction terminates for the programs MoA has actually been used to write, but whether it terminates for the full generality the $\lambda$-calculus pairing in Section VII.C would admit if taken at face value; resolving that gap, or arguing that MoA need not admit that full generality in the first place, is left as future work alongside orienting the identities into an explicit rewriting system. Resolving this --- orienting the identities into an explicit rewriting system, and then either proving termination for the resulting rules or stating precisely the class of expressions for which it is only observed --- is accordingly added to the open questions carried into the Conclusion.

\subsection{The Psi-Calculus Shares This Property}
The $\lambda$-calculus is Church--Rosser by the original 1936 theorem. The $\psi$-calculus --- the indexing calculus underlying MoA, built from $\psi$, $\iota$, and $\gamma$ --- is Church--Rosser as well: reducing a composed array/tensor expression to its DNF yields a unique result regardless of the order in which sub-expressions are $\psi$-reduced. This is stated directly in Mullin's own account of the Fortran realization of MoA \cite{fortran2024}: since both the lambda and psi calculi have the Church--Rosser property, proving the equivalence of array/tensor programs is possible. The same property is invoked in the DGEMM cache-blocking work \cite{thomas2021}. The earliest source for confluence of arrays under the $\lambda$-calculus is Klaus Berkling's 1990 Syracuse technical report, \emph{Arrays and the Lambda Calculus} \cite{berkling1990}; subsequent MoA literature, however, attributes the property to the Fengshui alignment paper \cite{chetioui2019} --- whose co-authors Abusdal, Haveraaen, and J\"arvi are at the University of Bergen, Norway --- rather than to Berkling directly, which suggests the more accessible or more directly worked argument may sit there rather than in the 1990 report. This article has traced the claim to both sources but has not yet obtained and checked either proof directly; a self-contained restatement of the argument, rather than a citation alone, is still needed for a fully rigorous treatment, and the Bergen paper is the more promising place to look first.

What confluence rules out, concretely, is worth checking against a small example rather than taken only on the strength of the citation above. Using the same $\xi$ from Section V.C, consider the composed expression $(\langle 0,1\rangle\,\psi\,\xi) + (\langle 1,2\rangle\,\psi\,\xi)$, summing two independently selected elements. The two $\psi$-selections do not depend on one another, so there are two orders in which a reduction could proceed. Reducing left to right: $\langle 0,1\rangle\,\psi\,\xi$ reduces first, to $\xi[0,1] = 11$; the expression becomes $11 + (\langle 1,2\rangle\,\psi\,\xi)$; the remaining selection reduces to $\xi[1,2] = 15$; addition gives $11 + 15 = 26$. Reducing right to left: $\langle 1,2\rangle\,\psi\,\xi$ reduces first, to $15$; the expression becomes $(\langle 0,1\rangle\,\psi\,\xi) + 15$; the remaining selection reduces to $11$; addition gives the same $11+15=26$. Both orders reach the same value by the same number of steps, which is unsurprising for an expression this small --- the substance of the confluence claim is that this keeps holding as expressions grow large enough, and are parallelized aggressively enough, that no person is checking every possible reduction order by hand, which is precisely the regime dimension lifting and multi-level $\gamma$ composition (Section VI) operate in.

\subsection{The Paradigm Claim: $\psi$-Calculus and $\lambda$-Calculus Together}
Backus's 1977 Turing Award lecture \cite{backus1978} argued for liberating programming from the ``von Neumann style'' via a variable-free algebra of programs. The suggestion that array programming specifically should be married to the $\lambda$-calculus predates MoA: Hai-Chen Tu and Alan Perlis, in \emph{FAC: A Functional APL Language} \cite{tu1986}, proposed a functional, $\lambda$-calculus-flavored treatment of APL and argued for the combination as a desirable direction for array programming generally. What Tu and Perlis's proposal did not, and could not, supply is the piece that makes the combination more than a stylistic preference: a $\psi$-calculus organizing indexing itself into an algebra confluent with the $\lambda$-calculus it is paired with, so that the guarantee of order-independence is a theorem about the pairing rather than an aspiration for it. Mullin's own retrospective states the connection between this line of thinking and MoA directly \cite{mullin1991}: MoA and the $\psi$-calculus were proposed as a route by which array programming could itself become a Formal Functional Programming (FFP) paradigm \emph{with} the $\lambda$-calculus, addressing anomalies left open in Iverson's array algebra and reaching the closure credited above to the 1988 dissertation --- supplying, in other words, exactly the missing piece Tu and Perlis's proposal had identified a need for but not itself provided. Because both calculi are confluent, a program built from $\lambda$-calculus function composition together with $\psi$-calculus array indexing inherits a single guarantee across both halves of the language: every well-formed expression, however it is reduced, reaches one and the same normal form. This is the precise, narrow sense in which the pairing can be called an \emph{ideal} paradigm for formal functional array programming and execution: not that it is convenient, but that it is the minimal formal combination in which neither function application nor array indexing can introduce order-dependent ambiguity into the result, from the abstract algebra all the way down to the address computed on physical hardware.

\section{Extending Pointer Arithmetic to the Mathematics of Arrays}
Section II noted that Yushchenko's Address Programming Language already permitted \emph{addresses of higher dimension} --- an address that itself contains an address, recursively, functionally equivalent to a pointer --- from its first variant in 1958. That capability has, until now, been described in this article only historically. It has a direct, and previously unstated, formalization inside the same $\psi/\iota/\gamma$ apparatus used throughout Sections V--VII, requiring no new primitive.

\begin{identity}[Indirection identity]
Let $\xi$ be an array of shape $\rho\xi$, and let $\eta$ be an array of shape $\rho\eta$ whose elements are themselves valid index vectors into $\xi$. Define indirect selection, for every index vector $\vec{p}$ in the index space of $\eta$, by
\[
(\eta \;\psi_{\mathrm{ind}}\; \xi)[\vec{p}] \;\equiv\; \big(\eta[\vec{p}]\big)\;\psi\;\xi \;\equiv\; (\rav\,\xi)\big[\gamma\big(\eta[\vec{p}] ; \rho\xi\big)\big] .
\]
\end{identity}
Since $\eta[\vec{p}]$ is itself obtained by $\psi$-selecting $\eta$ --- $\eta[\vec p] \equiv \vec p\,\psi\,\eta \equiv (\rav\,\eta)[\gamma(\vec p;\rho\eta)]$ --- indirect selection is exactly a \emph{composition} of two applications of the address-mapping identity (Identity 2): one $\gamma$ to locate $\vec p$'s entry in $\eta$, and a second $\gamma$ to locate the address that entry names in $\xi$. This is the precise algebraic content of Yushchenko's ``address of higher dimension'': not a new operator, but $\gamma$ composed with itself, $\gamma \circ \gamma$, one composition per level of indirection. An $n$-level pointer chain is $\gamma$ composed $n$ times.

This correspondence can be made exact rather than merely suggestive, by stating it directly in Yushchenko's own notation. Read $\eta$ in the role Yushchenko's set of addresses $A$ plays: an array each of whose elements is itself a valid index into some other array. Define the prime operation over $\eta$ by $'\vec p \equiv \eta[\vec p] \equiv (\rav\,\eta)[\gamma(\vec p;\rho\eta)]$ --- that is, Yushchenko's $'$ is exactly $\psi$-selection from $\eta$, restated in $\gamma$'s terms; this is the common semantics for the two notations that the correspondence requires. Under this reading, the second-dimension address Yushchenko writes $''\vec p_1 \equiv b$ becomes $\eta[\eta[\vec p_1]]$, two composed $\psi$-selections, which the identity above shows is exactly $\gamma$ composed with itself once. Yushchenko's construction of an address of the $n$-th dimension by $n$-fold application of $'$, and MoA's construction of an $n$-level pointer chain by $n$-fold composition of $\gamma$, are under this correspondence the same operation stated in two notations --- Yushchenko's over uninterpreted sets $A$ and $B$, MoA's over typed arrays and the shapes that index them --- rather than two operations that merely resemble one another.

Two consequences follow directly from results already established in this article. First, because the $\psi$-calculus is confluent (Section VII), a chain of indirections reduces to a unique address regardless of the order in which the constituent $\gamma$'s are evaluated --- an $n$-level pointer chase is as well-defined, and as safe to reorder or parallelize, as the single-level case in Identity~2. Second, this construction is recognizable as the array-language formalization of \emph{gather} and \emph{scatter}, the indexed load/store operations supported natively on vector hardware since the Cray X-MP: gather reads $y[i] \leftarrow x[\mathrm{idx}[i]]$ and scatter writes $y[\mathrm{idx}[i]] \leftarrow x[i]$, both special cases of the indirection identity above in which $\eta = \mathrm{idx}$. That gather/scatter already exists as a hardware primitive is, in the terms of this article, independent evidence that indirect addressing is not an awkward add-on to array algebra but a first-class citizen of it --- exactly as Yushchenko treated it from 1958, and exactly as dimension lifting (Section VI) already treats every other shape-to-hardware mapping in this framework.

This extension is offered here as a proposed formalization, not as a restatement of previously published MoA results; it has not yet been checked against an implementation or benchmarked, and per the Scope and Method commitments of this article it should be read as an open technical claim rather than an established one until that work is done.

\section{Why This History Matters: The Breadth and Stakes of the Mathematics of Arrays}
The throughline traced here --- Yushchenko's address, Iverson's whole-array notation, Abrams' shape-aware machine, Mullin's invention of indexing and closure, dimension lifting, and confluence --- is not a historical curiosity that terminates in one research group's tooling. The same $\psi/\iota/\gamma$ apparatus has been applied, published or currently in progress, to: radar signal processing \cite{mullin2005}; the Cooley--Tukey FFT \cite{cost2026}; dense linear algebra, where MoA-derived cache-blocking has been shown, in measured benchmarks, to exceed established DGEMM libraries \cite{thomas2021}; FPGA hardware co-design, where a combined inner/outer-product tensor algorithm was realized directly as digital hardware on a Xilinx Artix-7 FPGA \cite{grout2018}; GPU energy efficiency \cite{gpu2023}; GPU-targeted sparse linear algebra, where a MoA-derived operational normal form for sparse matrix--vector multiplication has been shown to maximize coalesced memory access on GPU hardware \cite{thomas2025gpu}; custom-architecture hardware-software co-design more broadly, where the FPGA realization above served as the experiment demonstrating the methodology itself, not merely an illustration of it \cite{grout2022}; standards-level language design, in ongoing work with the Fortran community --- principally with Arjen Markus --- toward MoA as a formally verified, enhanced subset of Fortran itself, rather than an external add-on \cite{fortran2024}; a parallel compiler effort connecting an enhanced subset of MoA formally to Python, carrying an array expression through DNF and ONF to a semantic subset of C by denotation-preserving algebraic rewrite --- a translator in the sense described above, not a compiler that must itself discover or verify an optimization --- rather than by trusting a compiler \cite{attention2026} --- itself the continuation of an earlier MoA-to-C-and-Fortran-90 compiler built at the University of Missouri--Rolla, documented in the Psi compiler project \cite{thibault1994} and in a concurrent undergraduate honors thesis on the same compiler, completed alongside Thibault's MS thesis \cite{umr1994honors}; emerging non-von-Neumann hardware, in current unpublished work with Nathaniel Cady on memristor crossbar processing-in-memory (PIM) architectures; and, most recently, transformer scaled dot-product attention \cite{attention2026}.

It is worth noting, as a check against the possibility that this is simply one research group citing itself, that the framework has been picked up as a reference point outside the immediate MoA group at least once: in the design of array-oriented database systems, where MoA is cited directly as prior art for algebraic reasoning about array layout and access \cite{vanballegooij2004}. This citation is now two decades old, and this article has not identified a more recent instance of independent uptake in the course of preparing it; the honest claim it supports is narrower than a pattern of adoption --- it is evidence that the formalism was, at least once, legible and useful to a researcher who did not develop it, not evidence of an ongoing independent research thread building on it. This is not a claim of large-scale adoption; MoA remains a specialist framework relative to mainstream array libraries such as NumPy or BLAS.

There is also a pedagogical case that has nothing to do with any single application. Most array-programming education proceeds by teaching a library's API and separately teaching performance folklore (cache blocking, memory alignment, vectorization) as an unrelated set of tricks acquired by experience. The lineage traced in this article offers a different proposition: that indexing, shape, and address computation are a single algebra, learnable as such, in which the performance folklore is not folklore at all but a theorem. Mullin's own framing of the ambition is direct: a formalism simple enough that any array-based function can be defined with shapes and $\psi$ alone, general enough to serve as a common substrate underneath the many array languages and libraries that currently have no shared theory connecting them.

\subsection{Toward a Universal Subset Principle}

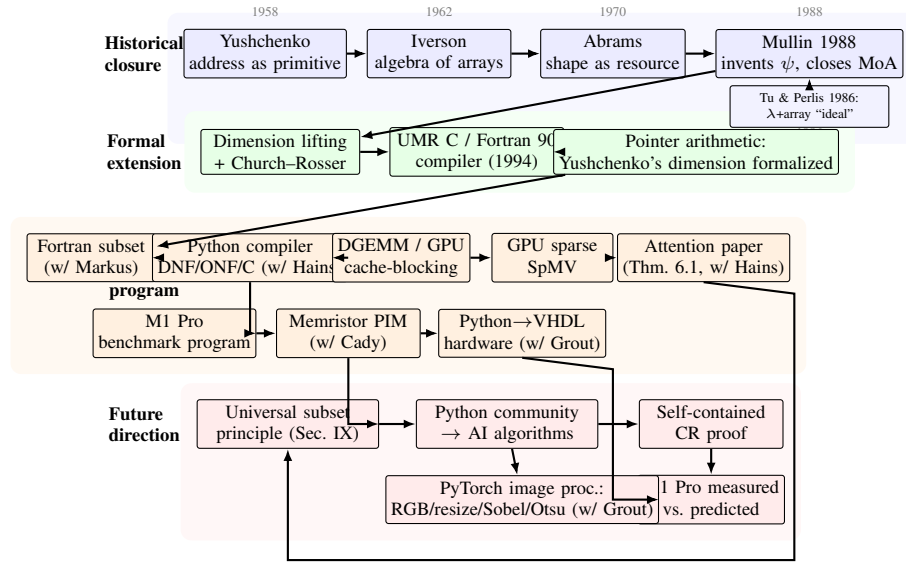
\begin{figure*}[t]
\centering
\begin{tikzpicture}[
  font=\scriptsize,
  box/.style={draw, rounded corners=1pt, align=center, minimum height=0.55cm, inner sep=2pt, fill=white},
  rowlabel/.style={font=\scriptsize\bfseries, align=left},
  arr/.style={-{Latex[length=1.6mm]}, thick},
  yr/.style={font=\tiny, align=center, text=gray}
]

\node[rowlabel] (l1) at (-3.3,4.6) {Historical\\closure};
\node[box, fill=blue!8, minimum width=1.9cm] (yush) at (-1.7,4.6) {Yushchenko\\address as primitive};
\node[box, fill=blue!8, minimum width=1.9cm] (iver) at (0.6,4.6)  {Iverson\\algebra of arrays};
\node[box, fill=blue!8, minimum width=1.9cm] (abr)  at (2.9,4.6)  {Abrams\\shape as resource};
\node[box, fill=blue!8, minimum width=2.3cm] (moa)  at (5.5,4.6)  {Mullin 1988\\invents $\psi$, closes MoA};
\draw[arr] (yush) -- (iver); \draw[arr] (iver) -- (abr); \draw[arr] (abr) -- (moa);
\node[yr] at (-1.7,5.15) {1958}; \node[yr] at (0.6,5.15) {1962};
\node[yr] at (2.9,5.15) {1970}; \node[yr] at (5.5,5.15) {1988};
\node[box, fill=blue!5, minimum width=2.1cm, font=\tiny] (tuperlis) at (5.5,3.9) {Tu \& Perlis 1986:\\$\lambda$+array ``ideal''};
\draw[arr] (tuperlis) -- (moa);
\node[yr] at (5.5,3.55) {1986};

\node[rowlabel] (l2) at (-3.3,3.3) {Formal\\extension};
\node[box, fill=green!10, minimum width=2.1cm] (dim) at (-1.5,3.3) {Dimension lifting\\+ Church--Rosser};
\node[box, fill=green!10, minimum width=2.3cm] (umr) at (1.1,3.3)  {UMR C / Fortran~90\\compiler (1994)};
\node[box, fill=green!10, minimum width=2.5cm] (ptr) at (4.0,3.3)  {Pointer arithmetic:\\Yushchenko's dimension formalized};
\draw[arr] (moa) -- (dim); \draw[arr] (dim) -- (umr); \draw[arr] (umr) -- (ptr);

\node[rowlabel] (l3) at (-3.3,1.6) {Current\\program};
\node[box, fill=orange!12, minimum width=1.7cm] (fort) at (-4.0,1.9) {Fortran subset\\(w/ Markus)};
\node[box, fill=orange!12, minimum width=1.9cm] (pyc)  at (-1.9,1.9) {Python compiler\\DNF/ONF/C (w/ Hains)};
\node[box, fill=orange!12, minimum width=1.8cm] (gemm) at (0.1,1.9) {DGEMM / GPU\\cache-blocking};
\node[box, fill=orange!12, minimum width=1.6cm] (spmv) at (2.1,1.9) {GPU sparse\\SpMV};
\node[box, fill=orange!12, minimum width=2.0cm] (attn) at (4.1,1.9) {Attention paper\\(Thm.\ 6.1, w/ Hains)};
\node[box, fill=orange!12, minimum width=1.7cm] (m1)   at (-2.9,0.9) {M1 Pro\\benchmark program};
\node[box, fill=orange!12, minimum width=1.9cm] (pim)  at (-0.6,0.9) {Memristor PIM\\(w/ Cady)};
\node[box, fill=orange!12, minimum width=2.1cm] (vhdl) at (1.7,0.9)  {Python$\to$VHDL\\hardware (w/ Grout)};
\draw[arr] (ptr) -- (fort); \draw[arr] (fort) -- (pyc); \draw[arr] (pyc) -- (gemm);
\draw[arr] (gemm) -- (spmv); \draw[arr] (spmv) -- (attn);
\draw[arr] (pyc) |- (m1); \draw[arr] (m1) -- (pim); \draw[arr] (pim) -- (vhdl);

\node[rowlabel] (l4) at (-3.3,-0.3) {Future\\direction};
\node[box, fill=red!8, minimum width=2.4cm] (univ) at (-1.4,-0.3) {Universal subset\\principle (Sec.\ IX)};
\node[box, fill=red!8, minimum width=2.4cm] (pyai) at (1.5,-0.3)  {Python community\\$\to$ AI algorithms};
\node[box, fill=red!8, minimum width=1.9cm] (crproof) at (4.2,-0.3) {Self-contained\\CR proof};
\node[box, fill=red!8, minimum width=1.9cm] (m1val) at (4.2,-1.3) {M1 Pro measured\\vs.\ predicted};
\node[box, fill=red!8, minimum width=2.3cm] (vision) at (1.7,-1.3) {PyTorch image proc.:\\RGB/resize/Sobel/Otsu (w/ Grout)};
\draw[arr] (attn.south) -- (5.3,1.425) -- (5.3,-2.1) -- (-1.4,-2.1) -- (univ.south);
\draw[arr] (pim) |- (univ); \draw[arr] (vhdl.south) -- (2.9,0.425) -- (2.9,-1.3) -- (vision.east);
\draw[arr] (univ) -- (pyai); \draw[arr] (pyai) -- (crproof); \draw[arr] (crproof) -- (m1val);
\draw[arr] (pyai) -- (vision);

\begin{scope}[on background layer]
\node[fit=(yush)(iver)(abr)(moa)(tuperlis), fill=blue!3, rounded corners, inner sep=6pt] {};
\node[fit=(dim)(umr)(ptr), fill=green!4, rounded corners, inner sep=6pt] {};
\node[fit=(fort)(pyc)(gemm)(spmv)(attn)(m1)(pim)(vhdl), fill=orange!5, rounded corners, inner sep=6pt] {};
\node[fit=(univ)(pyai)(crproof)(m1val)(vision), fill=red!3, rounded corners, inner sep=6pt] {};
\end{scope}
\end{tikzpicture}
\caption{The full arc this article traces, expanding the historical closure shown in Fig.~\ref{fig:addressing-evolution}: the historical arc traced in this article (blue), its formal extension (green), the current active research program (orange), and the direction it is heading (red). Boxes within a track run left to right in roughly chronological order; arrows crossing tracks show where present work grows directly out of the historical closure, and where future work grows out of the present.}
\label{fig:roadmap}
\end{figure*}

The applications listed above --- Fortran, Python, C, FPGA hardware description, and now memristor PIM --- can be read two ways. One reading treats them as a list: MoA has, so far, been connected to these particular languages and substrates, one at a time, by one research group's effort in each case. The other reading, and the one this article's ongoing research program is actually organized around, treats them as instances of a single underlying claim: \emph{any} interpreted, compiled, or otherwise executable language admits an MoA-conformant subset, and once the connection between that subset and the $\psi/\iota/\gamma/\rho/\rav$ apparatus is formulated for a given language, every optimization and correctness guarantee already established for MoA --- the DNF's architecture-independence, the ONF's address-optimality, dimension lifting's predictive instantiation, confluence's order-independence --- transfers to that language's subset without being re-derived from scratch.

What differs from language to language is not the primitive set itself, and this article does not treat that set as an open question. Mullin's 1988 dissertation defines every function in the algebra in terms of exactly five primitives --- $\psi$, $\iota$, $\rho$, $\rav$, and $\gamma$, per Section V --- with bracket selection as shorthand for $\psi$ rather than a sixth independent member of the set; any function that can be defined using shape ($\rho$) and the index generator ($\iota$) together with $\psi$ belongs in the algebra by construction, and is guaranteed, by the closure argument of Section V and the confluence property of Section VII, to compose correctly with every other function defined the same way --- composability is not checked function by function as new ones are added, it is inherited automatically from the closure of the primitive set itself. The per-language work is accordingly narrower than a search for which primitives a universal array language should provide: it is the single, purely correspondence-finding problem of identifying which of a host language's own existing constructs realize this already-fixed set of primitives, and stating that correspondence once, explicitly, rather than leaving it to be rediscovered informally by whoever next wants to apply MoA to a new language.

This reframing matters for where the current research program is actually going. The Fortran collaboration with Markus and the Python compiler work are not two separate case studies; they are two solutions to the same formulation problem, posed against two different host languages. The same is true of the historical C and Fortran~90 compiler at Missouri--Rolla, and of the current work with Nathaniel Cady connecting MoA's shape-to-hardware apparatus to memristor crossbar processing-in-memory (PIM) architectures --- a substrate with no instruction set and no conventional memory hierarchy at all, and therefore the sharpest test yet of whether the subset-formulation problem is genuinely language- and architecture-agnostic or only appeared so across the more conventional targets addressed so far. The widest span demonstrated so far, however, is the FPGA work with Ian Grout: a Python program formulated in MoA, taken through DNF and ONF, and realized as verified VHDL hardware, with the same translator --- not a separate hardware-design effort --- carrying it the entire distance from a high-level interpreted language to synthesized silicon. Grout and Mullin are currently extending this line directly into image processing, formulating RGB-to-greyscale conversion, resize, Sobel edge detection, and Otsu thresholding in Python and PyTorch as MoA expressions, with the same DNF/ONF/VHDL pipeline as the target --- a concrete, in-progress test of whether the universal subset principle holds for a class of AI-adjacent image-processing kernels rather than only for dense linear algebra. Stating the general principle explicitly, rather than leaving it implicit across a growing list of individually announced connections, is itself part of the unfinished work this article points toward: the formulation problem needs to be made easily accessible --- a stated procedure, not a bespoke research contribution required anew for every target --- before the claim that MoA subsets can be added to any language ``with all optimizations applied everywhere'' is more than an aspiration.

The nearest-term test of this is deliberately outward-facing rather than another internal case study: an active effort to engage the Python community directly, aimed at extending the existing Python/MoA compiler work into AI algorithm implementations specifically, where memory-bandwidth-bound kernels (attention among them, per Section VII) are exactly the case dimension lifting was built to address. Whether that engagement succeeds is an empirical question about community adoption, not a mathematical one, and this article does not overstate it: it is recorded here as the current direction of the research program, not as an accomplished result.

\section{Conclusion and Future Work}
This article has traced a single throughline in the history of computing --- the treatment of memory address as a first-class mathematical object --- from Yushchenko's Address Programming Language, whose first variant she proposed jointly with Korolyuk in 1958 out of a 1957--58 seminar, through Iverson's 1962 dimension-generalized algebra of whole-array operations without addresses or an indexing function, through Abrams' 1970 machine, which first put array shape to use as a resource for machine-level efficiency without yet formalizing an indexing function of any kind, to Mullin's 1988 invention of the indexing function itself and closure of composition of indexing, and onward through dimension lifting and the Church--Rosser property to a claim about predictive, provably correct instantiation on physical hardware.

What follows is stated in labeled parts rather than run together, matching the validated/theoretically-established/provisional discipline used throughout this article: one correction now closed, two gaps still open, and the future-work items that follow from each.

\emph{Correction closed.} The primary-source pass on Yushchenko, previously listed as an open gap, has now been carried out directly against her 1963 monograph. It corrected a specific error common to the secondary English-language record: the language originates from a 1957--58 seminar and a 1958 first variant developed jointly with V.~S.~Korolyuk, not from a solo 1955 invention as most retrospectives state.

\emph{Gap one: the confluence proof.} The Church--Rosser property claimed for the $\psi$-calculus is supported here by citation to both Berkling's 1990 report and the Bergen-affiliated Fengshui paper \cite{berkling1990,chetioui2019}, without a self-contained proof reproduced in this article. Checking the Fengshui paper first, since later MoA literature attributes the property to it directly, and reproducing whichever argument holds up, is a natural next step. As Section VII makes explicit, however, this is not simply a matter of transcribing an existing proof: confluence and termination presuppose an actual rewriting system, and MoA's identities have not yet been oriented into one anywhere in the literature, this article included. Orienting the identities is accordingly the logically prior step, on which both a confluence proof and a termination proof for $\psi$-reduction depend. The termination question is, however, narrower in practice than in principle: every application surveyed in this article is a primitive-recursive function, or a still less expressive one, and primitive-recursive functions terminate by construction. What remains open is not termination for MoA as actually used, but whether the full generality of the $\lambda$-calculus pairing claimed in Section VII.C is needed at all, or whether MoA's own applications are better served by a strictly weaker, always-terminating fragment; stating precisely the class of MoA expressions for which $\psi$-reduction is known to terminate, and resolving that broader question once the rewriting system is made explicit, is left as an open question rather than an assumed one.

\emph{Gap two: the empirical comparison.} The claim that MoA predicts an algorithm's ideal instantiation, rather than merely describing one post hoc, is validated for cache-blocked GEMM and GPU energy efficiency, and is now theoretically established for transformer attention as well, via a proved lower bound on data movement rather than an unproven conjecture \cite{attention2026}. What remains provisional is not that theoretical claim but its measured realization, pending the measured-versus-predicted comparison from the ongoing Apple M1 Pro benchmarking program and the planned ACCESS Accelerate HPC allocation.

\emph{Future work.} Four items follow directly from the gaps above and from Section IX's proposals, and are left to future work rather than assumed: extending the historical and technical account to the higher-dimension arrays required by transformer attention; testing the indirection identity proposed in Section VIII against a working implementation and benchmark; making the per-language correspondence procedure behind the universal subset principle stated in Section IX --- mapping a host language's own constructs onto the fixed $\psi/\iota/\rho/\rav/\gamma$ primitive set --- a stated, repeatable procedure rather than an aspiration; and testing that principle against the sharpest available case, memristor crossbar PIM, where no conventional instruction set or memory hierarchy exists to fall back on. 

\emph{What this article deliberately does not do.} It does not undertake a comparative survey of array-processing languages (later APL dialects, J, NESL, ZPL, and similar). The ongoing research program this history leads into is not aimed at that comparison, but at the abstract formulation itself and at practical engagement with communities already using array-based languages --- principally Fortran, via the collaboration with Markus, and Python, via the compiler work connecting an enhanced subset of MoA formally to it --- rather than at a genealogy of array languages for its own sake.

\bibliographystyle{IEEEtran}

\end{document}